# Visualization Optimization :
# Application to the RoboCup Rescue Domain

Pedro Miguel Moreira[1,4,] Luís Paulo Reis[2,4] and A. Augusto de Sousa[3,4]

1ESTG-IPVC - Escola Superior de Tecnologia e Gestão de Viana do Castelo, Portugal
2LIACC - Laboratório de Inteligência Artificial e Ciência de Computadores da Univ. Porto, Portugal
3INESC-Porto - Instituto de Engenharia de Sistemas e Computadores do Porto, Portugal
4FEUP - Faculdade de Engenharia da Universidade do Porto, Portugal

**Abstract**

*In this paper we demonstrate the use of intelligent optimization methodologies on the visualization optimization of virtual / simulated environments. The problem of automatic selection of an optimized set of views, which better describes an on-going simulation over a virtual environment is addressed in the context of the RoboCup Rescue Simulation domain. A generic architecture for optimization is proposed and described. We outline the possible extensions of this architecture and argue on how several problems within the fields of Interactive Rendering and Visualization can benefit from it.*

**Additional Info**

corresponding author : pmoreira@estg.ipvc.pt

# Visualization Optimization : Application to the RoboCup Rescue Domain

Pedro Miguel Moreira[1,4†], Luís Paulo Reis[2,4‡] and A. Augusto de Sousa[3,4§]

[1]ESTG-IPVC - Escola Superior de Tecnologia e Gestão de Viana do Castelo, Portugal
[2]LIACC - Laboratório de Inteligência Artificial e Ciência de Computadores da Univ. Porto, Portugal
[3]INESC-Porto - Instituto de Engenharia de Sistemas e Computadores do Porto, Portugal
[4]FEUP - Faculdade de Engenharia da Universidade do Porto, Portugal

**Abstract**

*In this paper we demonstrate the use of intelligent optimization methodologies on the visualization optimization of virtual / simulated environments. The problem of automatic selection of an optimized set of views, which better describes an on-going simulation over a virtual environment is addressed in the context of the RoboCup Rescue Simulation domain. A generic architecture for optimization is proposed and described. We outline the possible extensions of this architecture and argue on how several problems within the fields of Interactive Rendering and Visualization can benefit from it.*

Categories and Subject Descriptors (according to ACM CCS):  I.3.7 [Computer Graphics]: I.3.7 Virtual reality I.2.8 [Artificial Intelligence]: Problem Solving, Control Methods, and Search

## 1. Introduction

In this paper we address the problem of automatically find a fixed set of views (multi-view) over a three dimensional dynamical and evolving simulated environment that gives to the user a good representation. We report the application to the RoboCup Rescue Domain where we aim at obtaining, at each moment, the set of views which better describe the emergency situations and rescue operations. These multi-view should provide the user with useful information leading to a correct understanding of the whole environment.

The problem of finding the best set of views over a scene can be stated as an optimization problem. This is an interesting and useful problem which relates to other problems such as : automatic object [VS03] and scene exploration [AVF04], virtual camera motion  [MC00], virtual cinematography [DZ95] and automatic selection of images to Image-Based Modeling and Rendering (IBM&R) systems [VFSH03]. Finding the optimal solution is not usually feasible due to the inherent problem complexity and to time constraints. Thus, meta-heuristics are used to find a best (suboptimal) solution.

In our approach we propose an optimization architecture relying on an optimization agent that works autonomously from the main visualization / rendering application. Our objective is to develop the proposed architecture in order to apply it on other problems with minor effort.

The optimization agent provides a set use of intelligent optimization techniques [PK00], such as (but not restricted to): genetic algorithms, tabu search, simulated annealing, or neural networks to efficiently find optimized solutions.

The rest of the paper is organized as follows. Following this introduction, Section 2 describes the aplication domain - Robocup Rescue, where reported experiments were conducted. Next, in Section 3, we detail and formalize the addressed problem. Our proposed methodology is presented in Section 4. In Section 5 experimental results are presented and discussed. Finally, in Section 6, conclusions and outlines of our on going and future work are presented.

† pmoreira@estg.ipvc.pt
‡ lpreis@fe.up.pt
§ augusto.sousa@fe.up.pt

○

## 2. RoboCup rescue domain

RoboCup was created as an international research and education initiative, aiming to foster artificial intelligence and robotics research, by providing a standard problem, where a wide range of technologies can be examined and integrated [Ano06].

The huge success of the RoboCupSoccer international research and education initiative, led the RoboCup Federation to create the RoboCupRescue project focussing on Urban Search and Rescue (USAR) operations [Ano06].

The RoboCupRescue Simulation League consists of a simulated city in which heterogeneous simulated robots, acting in a dynamic environment, coordinate efforts to save people and property. Heterogeneous robots in a multi-robot system share a common goal, but have different abilities and specializations, adding further complexity and strategic options. These systems can manifest self-organization and complex behaviors even when the individual strategies of all the robots are simple. The team-programmed robots are of three different types: Fire Brigades, Police Forces and Ambulance Teams. Fire Brigades are responsible for extinguishing fires; Police Forces open up blocked routes; and Ambulance Teams unbury Civilians trapped under debris. Each of these types of robots is coordinated by an intelligent centre responsible for communication and strategies. In order to obtain a good score, all these robots work together to explore the city, extinguish fires, and unbury Civilians.

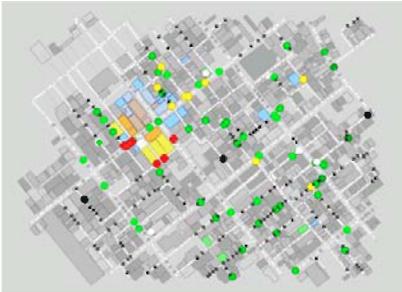

**Figure 1:** *RoboCup Rescue Simulated Environment*

The simulated environment, Fig. 1 is composed of several sub-simulator modules such as the fire simulator and blockade simulator. This structure allows for independent module development, permitting the addition of new modules and making the simulator system more realistic.

## 3. Visualization of RoboCup Rescue Simulations

There are several tools for visualizing RoboCup Rescue Simulations. Log viewers are used to track the evolution and result of rescue simulations. Some viewers have been written so far, enabling different viewing perspectives of the simulation, but all of them lack the functionality of a good debugging viewer [Ari05]. Some teams have performed work in this area, but the need for a more comprehensive tool was only made more visible [Ari05]. The Freiburg team, amongst others, has developed its own viewer, releasing it to the rescue community. Freiburg's 3D viewer [KG05] is one of the most used by the community, second only to Morimoto's 2D viewer [Mor02], which is included in the official simulator package.

Our purpose was to develop a visualization tool to the Robocup Rescue Domain, that features a multi-view over the simulated environment. Camera positions are restricted to existing rescue agents or entities (such as buildings, police, fire brigades, etc). Aerial views are planned but not yet implemented. Users monitoring the rescue simulation should benefit from such tool since they are provided with a fixed (and small, e.g. four) number of views selected based on criteria that tries to optimize the relevance of the virtually captured imagery to the understanding of the evolving simulation. Our viewer is partially based on [KG05].

### 3.1. Problem description

The problem can be informally stated as: In an urban rescue setting there are $m$ visualization agents that can obtain views over the scene. The objective is to find an optimal set of k views that better describe the simulation for each moment. These $k$ views can have a different purpose, but for the sake of simplicity, we restrict the discussion to the monitoring of emergency situations. The visualization agents are controllable in the sense that one can affect their viewing parameters.

The optimization problem can be formalized as follows:

$E = \{e_1, \ldots e_n\}$

$V = \{v_1, \ldots v_m\}$

$v_i = f(\vec{Pos}_i, \vec{VD}_i, \vec{VUP}_i, FoV_i)\ i \in \{1, \ldots, m\}$

$MV = \{mv_1, \ldots, mv_k\}$ where $MV \subset V$ and $mv_i \neq mv_j\ \forall i \neq j$

MAXIMIZE: $Q(MV) =$
$\sum_{j=1}^{k} \sum_{i=1}^{n} Vis(e_i^j).Red(e_i|MV)(W_1.Rel(e_i) + W_2.Ecc(e_i^j))$

where, $E$ is the set of $n$ entities that have relevance in the scene (buildings, agents, etc). $V$ is the set of different views (equals the number of agents/entities with viewing capabilities). Each view is characterized by common camera parameters, as the position $\vec{Pos}_i$, view direction $\vec{VD}_i$, relative camera orientation $\vec{VUP}_i$ and field of view $FoV_i$. Aspect ratio is not being considered as it remains unchaged. $MV$ is a multiview setting consisting of $k$ distinct views from $V$. The problem is to find the optimal $MV$ set, with appropriate view parameters, that describes the rescue scenario with

○

better quality. Quality $Q$ is computed using the following criteria (note that $e_i^j$ denotes the visual properties of an entity $e_i$ in a image obtained by the view $j$)

**Visibility:** $Vis(e_j^i)$ Only visible objects are considered to the solution. For those they can be displayed with more or less pixels on the image due to their visible projected area. The visible projected area relates to the object's distance to the viewer, to its size and orientation, and also to how much partial occlusion it suffers from other objects.

**Relevance:** $Rel(e_i)$ A measure of how relevant is that object for the purpose of the visualization. For example, if tracking emergency situations, a building on fire will have a greater relevance than an unaffected building. The intrinsic importance of an object is also considered, e.g. hospitals, fireman headquarters, schools have more relevance than ordinary buildings.

**Redundancy:** $Red(e_i|MV)$ It is expected that the multiple set of views describe as much as possible distinct situations occurring during the simulation. Thus, redundant views over the same objects are penalized.

**Eccentricity:** $Ecc(e_j^i)$ A measure on how distant to the center of the image an object is displayed. There is a penalty for more eccentric objects based on the fact that users pay more attention to central objects in the image.

## 4. Optimization Methodology

We propose an architecture relying on an optimization agent that operates detached from the main application. As showed in Fig. 2, communication is achieved by means of a simple protocol (in the figure a simplified version is depicted). The optimization agent (OA) informs the visualization application (VA) that it is available with a connect message. The VA acknowledges the connection and sends a problem description. The OA requests data relevant to the optimization process. As optimized solutions are computed, they are communicated to the VA which sets up and operates appropriately.

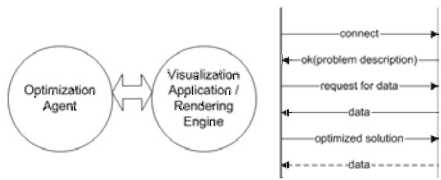

**Figure 2:** *The optimization architecture and a simplified version of the communication protocol*

Further communication between the two processes should be well established by means of a proper language. Some desirable features of such a language are:

- Independency of implementation details;
- Small but easily extensible;
- Generality of its format and possible incorporation of different abstraction levels;
- Robustness, with simple data validation and override of usual errors.

The main advantages of this architecture are:

**Generality** The architecture is designed to be easily applied to a wide variety of problems.
**Flexibility** Its behavior should adapt to the demands of the visualization application.
**Modularity** The design of the supporting architecture enables its extension by modular pieces.
**Portability** Operating at a higher level and being independent from the implementation details of the visualization application and the rendering engine it is expected to be easily portable or used by many different systems.

As a main drawback there is some possible loss of performance comparatively to "proprietary" approaches were optimization is hardly embedded within the visualization processes or the rendering pipeline. This loss of performance is fundamentally due to the necessary communication overhead introduced by the detachment of such processes.

## 5. Experimental results

In order to evaluate usefulness of our approach, we are implementing the proposed architecture and conducting our first experiments. A scenario with 1035 relevant entities and 50 agents with viewing capabilities was used.

As it is expected that the quality function has several local optima, simulated annealing [KGV83] (SA) was chosen as the meta heuristic as it has the ability to continue the search even if a local optimum is found. Another contributing reason is its computational efficiency compared to other meta-heuristics(e.g. genetic algorithms). At each iteration SA considers a neighbour of the current state, and probabilistically decides on moving to it or staying in the current state. The probabilities are chosen so that the system ultimately tends to move to states with better quality.

Our neighbour states are obtained, by exchanging one view from the multiview set or by changing one of the view parameters. In this experiment we are not changing the view position, but using the agent's position during the rescue simulation. We are also exploring the concept of adaptive neighborhood by adaptively defining the range of change in view parameters as a function of the evolution of quality. Note that for problems where the conditions vary with some kind of continuity, the optimization step (as well as visibility determination) can benefit from the exploration of spatial and temporal coherence.

In Fig 3 the evolution of quality can be observed for an optimization of a multivew consisting in four views. The solution converged to a stable maximum on approximately 500 iterations.

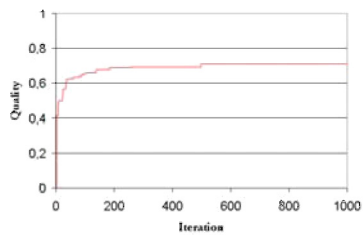

**Figure 3:** *Evolution of the quality of the views along the optimization process.*

The exact visible set is necessary to determine whether an object is actually visible. We are using pre-processed from-region visibility and then using an item-buffer technique for exact visibility. Image analysis reveals the visible objects as well as coverage statistics. Histogram utilities from OpenGL can be used to improve performance. Further acceleration can be achieved if considered that objects with small visible areas are not relevant. In such case the above process can be done at smaller image resolutions. We are also investigating the use of OpenGL hardware occlusion queries. Eccentricity is estimated by projecting the center of the objects bounding boxes.

## 6. Conclusions and Future Work

We have presented and described an optimization architecture, suitable to be applied to several visualization and rendering problems. We pointed several advantages of our approach.

In our first experiments, we have applied this architecture to a visualization problem in the Robocup rescue domain. Our first results are promising and demonstrate its usefulness.

As future work we plan to define a framework based on the proposed architecture including several optimization methods. Also planned is the formal specification of a language for efficient communication between the optimization agent and the rendering /visualization application.

Another guideline of our research concerns to path advice leading to assisted exploration of scenes. Due to high model complexity or scene extension, unguided exploration by the user can result on poor explorations, namely by missing relevant parts of the model. The expected result will be a system that generates and suggests to the user optimized paths according to the users interests, advertisements, etc.

The proposed architecture could be also extended and adapted to be used in other areas, namely in remote rendering [TL01]. Optimizing the quality of the visualization in remote rendering architectures depends on many factors. Examples of such factors are the rendering potential of the client and the available bandwidth. Deciding on when and which information should be transmitted to the client and on how to balance the rendering load between the server and the client is a difficult problem that we think can substantially benefit from intelligent optimization strategies.

## Acknowledgements

The first author acknowledges the support by European Social Fund program, public contest 1/5.3/PRODEP/2003, financing request no. 1012.012, medida 5/acção 5.3 - Formação Avançada de Docentes do Ensino Superior, submitted by Escola Superior de Tecnologia e Gestão do Instituto Politécnico de Viana do Castelo. This work was partially supported by the project *Rescue: Coordination of Heterogeneous Teams in Search and Rescue Scenarios* - FCT/POSI/EIA/63240/2004.